# Integrated Data-Driven Biotechnology Research Environments

Rosalia Moreddu[1,2,*]

[1]School of Electronics and Computer Science, University of Southampton, Southampton, UK

[2]Institute for Life Sciences, University of Southampton, Southampton, UK

*r.moreddu@soton.ac.uk

## Abstract

In the past few decades, the life sciences have experienced an unprecedented accumulation of data, ranging from genomic sequences and proteomic profiles to heavy-content imaging, clinical assays, and commercial biological products for research. Traditional static databases have been invaluable in providing standardized and structured information. However, they fall short when it comes to facilitating exploratory data interrogation, real-time query, multidimensional comparison and dynamic visualization. Integrated data-driven research environments aiming at supporting user-driven data queries and visualization offer promising new avenues for making the best use of the vast and heterogeneous data streams collected in biological research. This article discusses the potential of interactive and integrated frameworks, highlighting the importance of implementing this model in biotechnology research, while going through the state-of-the-art in database design, technical choices behind modern data management systems, and emerging needs in multidisciplinary research. Special attention is given to data interrogation strategies, user interface design, and comparative analysis capabilities, along with challenges such as data standardization and scalability in data-heavy applications. Conceptual features for developing interactive data environments along diverse life science domains are then presented in the user case of cell line selection for *in vitro* research to bridge the gap between research data generation, actionable biological insight, experimental design, and clinical relevance.

## 1. Biological data management

Biology, nanotechnology and medicine are data-rich fields.[1] Over the last several decades, high-throughput technologies have revolutionized biology by generating massive datasets. These include genomic sequences, proteomics data, high-resolution imaging, long-term acquisitions, and clinical trial data.[2] On top of those, companies in the biotech industry have commercialized large amounts of biological models to be used in research, biotechnology, and pharmaceutical industries for *in vitro* research.[3] In response, the need for versatile and user-friendly resource and data management systems has grown dramatically.[4] Biological databases traditionally focused on cataloguing discrete pieces of information and statically showing them online (e.g., Cellosaurus for classifying cell lines)[5] or within private organizations (e.g., internal databases for storing laboratory equipment information). In some cases, they integrate simple search functions to facilitate retrieval of stored data and allow incremental data submission or periodic expansion by database curators.[4] Classic examples like GenBank[6] and the Protein Data Bank (PDB) offer comprehensive search and retrieval functions across standardized metadata fields (e.g., organism, gene name, accession number, sequence features in GenBank). Such systems remain indispensable as reference sources, but they were largely designed to support data deposition, retrieval and preservation, rather than interactive exploration or adaptive reuse. Their architecture typically centers around rigid schemas with limited user-driven comparison capabilities.

With the advent of high-throughput technologies, the volume and complexity of biological data expanded considerably.[7] In domains such as genomics, drug discovery, *in vitro* research and personalized medicine, interactive and integrated platforms have the potential to transform the way we work with data, reducing time currently devolved to hypothesis testing and literature search, and facilitating discovery by designing meaningful workflows based on experimental objectives. This model would also enable scientists to focus their efforts on innovation and higher-end intellectual activities. **Table 1** presents the comparison between traditional biological databases and the proposed approach. **Figure 1** visualizes the potential of interactive and integrated data environments. The following sections guide the development of next-generation digital research platforms.

The concept of interactivity in database systems has been widely used to describe data portals or repositories with web or API access. Here, this concept is expanded to include dynamic, modular systems designed for bidirectional interaction, collaborative filtering, hypothesis generation, experimental planning, and feedback integration. This allows to incorporate experimental metadata, support multiscale comparative analysis, and integrate FAIR Digital Objects derived from user interaction. Although no universally accepted term yet exists for such environments, this conceptual framework lies at the intersection of intelligent decision-support platforms, collaborative data infrastructures and multi-domain experimental design engines.

## 2. Interactive Data Environments

The concept of interactivity in database systems is closely related to developments in web technologies, artificial intelligence (AI), and data visualization techniques. Modern systems are capable to combine web technologies (for example JavaScript libraries for dynamic visualization)[8] and back-end data management solutions (e.g., NoSQL databases for unstructured data or graph databases for relationship modeling).[9] Integrated research environments are not intended to replace domain-specific repositories like GenBank, PDB, or Cellosaurus, which remain foundational for primary data submission and standardized archival. Instead, they could function as specialized integration and analysis layers that enable researchers to query, visualize, and analyze data across multiple existing repositories through unified interfaces. This architectural approach shares conceptual similarities with data lakes,[10-12] yet extends beyond traditional data lake implementations in several crucial ways. Data lakes primarily provide infrastructure for storing heterogeneous data in native formats without imposing rigid schemas,[13] whereas the research environments described here focus on active knowledge integration and experimental decision support through domain-specific analytical capabilities. Recent implementations such as genetic data lakes for drug discovery[10] store and process genetic data at scale, but typically lack the specialized comparative analysis capabilities and experimental design guidance that define the systems proposed hereby. General-purpose data lakes prioritize accommodating massive volumes of raw data,[14] these biotechnology research environments implement domain-specific user interfaces and analytical workflows optimized for particular scientific tasks (e.g., cell line selection, pathway analysis, biomarker identification). Furthermore, while data lakes typically operate as centralized repositories within organizational boundaries,[15] the research environments envisioned here function as connection across the distributed ecosystem of existing biological repositories. They provide harmonized access layers that preserve the specialized governance and data submission workflows of underlying repositories while enabling cross-repository analyses not feasible through direct interaction with individual primary data sources.

Despite the dramatic development in computer science and web technologies, the life science domain still sees crucial gaps to enable smooth selection and dataset navigation.[16, 17] The need of transitioning toward these features is becoming evident through the growing complexity of biological questions, in parallel with the technological advances in other fields that make complex computations and visualizations feasible in real time and with less efforts from the user.[18, 19] The next subsections highlight selected desirable characteristics and their technological feasibility within interactive data environments for the life sciences domain. Cell line selection for *in vitro* research is presented as a possible implementation case.

### 2.1. Features

The heterogeneity of biological data, from structured clinical trial tables and semi-structured cell line annotations to unstructured experimental notes, poses a fundamental challenge for the development of integrated data environments.[20] Successfully integrating these different data types requires a strategic balance to ensure that the system accommodates evolving data landscapes without sacrificing analytical precision. At the core of this integration lies the concept of adaptive data modeling (ADP),[21] where the choice of database schemes dictates both functionality and scalability. Relational models with a rigid table structure are indispensable for managing structured data like genomic variants, patient demographics, cell lines properties.[22] However, the dynamic nature of life sciences research sometimes demands schemeless architectures. In this context, document-oriented databases (for example MongoDB)[23] could be employed, allowing nested structures to capture variable data, for instance that associated with single-cell sequencing experiments.[24]

For highly interconnected data, such as protein-protein interaction networks or metabolic pathways in cells, graph databases (e.g., Neo4j) offer the required traversal speed to enable real-time queries across millions of nodes and edges.[25]

Data pipelines require automated workflows that analyze raw FASTQ files,[26] screen online publications for experimental conditions, or obtain real-time sensor data from laboratory equipment.[27, 28] Tools with error-handling frameworks could standardize this process, for example Apache NiFi or custom Python scripts, but challenges exist.[29] For instance, inconsistencies in how labs report cell line contamination status require context-aware natural language processing (NLP) models to normalize inputs.[30, 31] However, these models themselves may introduce noise or bias, especially when trained on incomplete or poorly annotated datasets. In addition to inconsistencies, metadata may be entirely missing or provided in minimal form, despite repository guidelines requesting rich contextual descriptors. Submitters may also inadvertently provide erroneous information due to lab tracking errors or manual entry mistakes. These factors further complicate data harmonization and highlight the need for robust validation mechanisms, error propagation awareness, and contributor-facing feedback loops within interactive platforms.

To bridge this gap, hybrid interfaces are gaining traction, for example Galaxy Project combining drag-and-drop workflows with Python scripting to allow users to transition from structured prompts (e.g., "show all breast cancer cell lines in the database having HER2+ status") to programmatic analyses (e.g., suitability analysis for a given experiment evaluated using R libraries).[32] Then, dynamic visualization could transform these raw query results into actionable insights. Modern systems integrate libraries to render interactive plots for genome-wide association studies, optimal cell profile to test a given technology, or 3D protein structures visualization. Examples of employed libraries are Plotly or D3.js.[33] However, interactivity implies that visualization must extend beyond passive observation. Interactive data environments could let users click on a graph or comparison plot to trigger a secondary query, for example extract all genes differentially expressed in a specific cluster, or group up all cells in different subclusters based on user-prompted features. Coupling visualization with analytical tools could enable this functionality, and assistive AI could amplify this interactivity and scope. All these functions drive the design of a suitable user interface (UI) and define user experience (UX) features.

### 2.2. Case Study: Biological Cell Line Selection

Cell line selection represents an ideal case study for demonstrating the potential of Interactive data environments in the life sciences. The complexity of choosing appropriate cell models from the thousands or commercially available lineages exemplifies why traditional static repositories are insufficient for modern research needs.[34] Currently, researchers often select cell lines based on convenience, tradition, or limited familiarity rather than comprehensive biological relevance, leading to potential experimental irreproducibility, translational failures, and wasted resources.[34, 35] The challenge lies in navigating multidimensional considerations simultaneously, spanning from genetic background, tissue origin, disease relevance, authentication status, growth characteristics, pathway activations, compatibility with experimental procedures, and more.[34] While valuable reference resources exist (Cellosaurus, ATCC catalogs, LINCS),[36] they typically present information in isolation, making comparative analysis labor-intensive and prone to oversight of critical variables if run by humans.

An Interactive data environment would transform cell line selection by enabling researchers to dynamically filter, compare, and visualize multiple cell lines across diverse parameters simultaneously. Such a system would integrate disparate data sources, including existing static repositories, literature outcomes, genomic profiles and user-contributed experimental metadata, creating a thoughtful support platform. The following subsections examine a potential envisioned architecture for implementation, structured into three interconnected layers: back-end data management systems for storing and processing diverse cell data, middleware and APIs facilitating integration and communication, and front-end technologies enabling intuitive exploration, comparison and visualization. This is also schematized in **Figure 2**.

#### 2.2.1. Back-End Data Management Systems

The back-end infrastructure forms the pillar of any interactive data environments[37, 38] where storage solutions are selected based on the inherent structure of biological information. Traditional relational database management systems (RDBMS) like PostgreSQL and MySQL build the foundations of many established biological repositories.[39] An example is Cellosaurus, a comprehensive repository of cell lines.[5] While invaluable as a reference, its traditional structure limits the utilization of the stored data. Currently, data can be visualized one by one for each cell line, making multidimensional comparisons across tens or hundreds of cells unfeasible. Document-oriented NoSQL databases offer significant advantages for cell line repositories that accumulate diverse experimental metadata.[40] MongoDB, for instance, can store cell lines as flexible JSON documents.[23] This allows to incorporate new characterization data of various nature, from morphological features to authentication profiles, without disruptive changes in the fundamental structure.

Network-oriented biological data presents another storage challenge that graph databases address. Systems like Neo4j could transform the way relationships between cell lines are accessed and understood.[25] In cancer research, one could explore the connections between patient-derived xenografts, immortalized cell lines and original tumor samples through intuitive graphs and user-driven multidimensional queries. This approach could reveal lineage relationships and experimental compatibility that remain obscured in conventional tabular repositories. Supplementary storage technologies could then drive performance considerations. Response times during comparative analyses could be dramatically reduced by employing key-value stores like Redis to catch frequently accessed data, for example commonly requested cell lines or culture protocols. This hybrid storage approach would allow databases to maintain responsive performance even as users perform complex multiline comparisons simultaneously.

### 2.2.2. Middleware and APIs

The middleware layer in a database typically orchestrates communication between storage systems and user applications,[41] in this case enabling to transform static cell line references into dynamic research tools. Instead of making separate requests for each cell line of interest, queries that directly compare multiple lines across selected parameters (growth kinetics, drug sensitivities, genetic backgrounds) in a single operation could be constructed. Examples of tools to achieve this include GraphQL over traditional RESTful APIs.[42] AI represents one of the most powerful middleware integrations.[43] Machine learning microservices could analyze patterns across thousands of cell lines to recommend optimal models for specific research questions. Drug screening experiments could be informed by recommendation assistants that identifies cell lines most relevant to their target pathway based on expression profiles, previous experimental outcomes, and literature associations. Such systems transform passive cell line catalogues into active research planning tools, integrating them with the latest research findings. The latter could be, in turn, standardized over time by researchers themselves who engage with the Interactive data environments.

The microservices architectural pattern can partition monolithic applications into independent, specialized components, enhancing system flexibility.[44] A modernized cell line database should separate authentication verification, experimental condition optimization, and cross-reference resolution into discrete services. When researchers upload new characterization data for a cell line, a validation microservice could automatically verify consistency with existing profiles, while another service updates recommended culture conditions based on combined experimental outcomes. Another key middleware component is workflow services, including laboratory information management systems (LIMS),[45] facilitating the link between information and action. An example is comparing metabolic profiles across hepatocyte cell lines to generate customized experimental protocols based on optimal culture conditions for each line, with reagent lists automatically adjusted for the specific metabolic properties of selected models.

### 2.2.3. Hybrid Architecture and Data Integration Framework

The proposed database architecture adopts a hybrid model combining a centralized repository for high-frequency primary data (e.g., cell line identifiers, validated traits) with federated integration of distributed secondary sources (e.g., genomic and literature databases) via standardized APIs. A three-tiered mediation layer enables technical integration. The physical access layer could employ OAI-PMH and bioinformatics standards such as GA4GH to standardize data harvesting. The middleware layer performs schema mapping for structured data and semantic normalization for unstructured content. At the top, a global identifier registry utilizes CURIEs and DOIs for cross-referencing across heterogeneous resources.
Three core interfaces are meant to ensure interoperability: (1) a unified query interface translating user requests into native database languages (SQL, Cypher, MongoDB);

(2) a synchronization interface maintaining consistency across relational, document, and graph stores; and (3) a metadata exchange interface compliant with ISA-Tab for harmonizing biological experiment descriptions. These enable referential integrity across datasets and pathway graphs. A feedback integration pathway captures user interactions (e.g., annotations, experimental suitability tags) as FAIR Digital Objects with standardized metadata. A dedicated ingestion module validates and reintegrates these into the core database, creating a dynamic knowledge refinement loop. This approach might allow the system to evolve collaboratively, integrating expert input and computational insights to enhance both analytical functionality and data richness over time.

### 2.2.4. Front-End Technologies

The front-end layer transforms static cell line catalogs into dynamic research platforms through intuitive interfaces for users.[46] This layer harmonically presents the features of the previous two layers through visualization and interactive features. Modern JavaScript frameworks enable sophisticated comparative visualization and interaction.[46] React components could render comparisons of cell lines, highlighting differences in morphology, growth characteristics, gene expression profiles, clinical relevance, usage, and much more, through interactive visualizations that respond instantly to parameter adjustments driven by the user. Visual comparison tools represent a simple yet potentially game-changing tool in cell line selection. Interactive matrices could display drug sensitivity patterns across multiple cell lines, with hierarchical clustering revealing unexpected relationships between seemingly unrelated models. This cross-check at multiple levels would likely produce precious novel insights based on fully exploiting and interpreting already-existing data. These comparisons could be filtered based on the most disparate parameters, needs or curiosities, based on specific mutations, tissue origins, or experimental conditions, transforming what would be weeks of literature review into minutes of interactive exploration.[34]

AI assistants integrated directly into the front-end interface could provide tailored and specifically trained guidance through cell line selection processes based on experimental goals.[47] For example, they could highlight cell lines with relevant properties, flag potential authentication concerns, contradictory experimental findings from the literature or suggest complementary models to strengthen experimental design. Finally, virtual cell culture simulators would be a transformative front-end addition to integrate mathematical models of cell behavior with accumulated experimental data. This could facilitate the prediction around how different cell lines might respond to experimental manipulations before physical experiments begin, enabling to adjust culture conditions, test drug concentrations, or simulate time-course experiments through intuitive interfaces, with predictions based on historical data. To support collaborative data exploration and hypothesis generation, the front-end interface is designed to enable real-time multi-user interaction. This would enable to synchronously examine data views, annotate selections, and share insights within team environments. Examples of technologies are Firebase Realtime Database or Socket.IO, employed in collaborative sessions where selections and filters are mirrored across team members interfaces in real time. This setup might facilitate distributed but coordinated interpretation of complex biological datasets.

## 3. Selected Applications of Dynamic Data Software

The applications of Interactive data environments span the entire spectrum of life sciences. In domains where relationships between entities are multidimensional and contextual, static tabular presentations fail to capture the complexity of biological systems. Interactive data environments are meant to integrate both structured and unstructured data. While structured data (genomic variants, protein structures, clinical measurements) forms the foundation, unstructured data (scientific literature, clinical notes, experimental protocols) provides crucial context at a given time. The following subsections examine four domains where Interactive data environments are demonstrating and can further have particular impact: genomics, drug discovery, systems biology, and clinical research. Each case explores how the interactive paradigm can address domain-specific challenges and transforms research practices.

### 3.1. Genomics

The genomics field has been an early adopter of Interactive data environments approaches, driven by the complexity and volume of sequencing data which made this route inevitable.[48] Genome Aggregation Database (gnomAD) evolved from simple variant browsers to sophisticated interactive systems to explore allele frequencies across populations, visualize genomic contexts, and assess functional impacts of variants in real-time.[48] These capabilities have proven to be critical for rare disease research. Modern genomic Interactive data environments integrate machine learning algorithms that predict variant pathogenicity while allowing users to adjust parameters based on domain knowledge.[49] For example, the ClinGen Pathogenicity Calculator enables clinicians to interactively apply ACMG guidelines for variant classification while visualizing supporting evidence from multiple sources.[49] This represents a significant advance over static variant lists, enabling interpretation that adapts to evolving clinical knowledge.

### 3.2. Drug Development

In pharmaceutical research, interactive data environments are revolutionizing multiple stages of the drug discovery and development pipeline. DrugBank systems allow to explore drug-target interactions across chemical and biological space.[50] Modern implementations combine structural databases with molecular docking algorithms, allowing users to interactively modify potential compounds and visualize predicted binding affinities in real-time. Virtual screening applications particularly benefit from interactive capabilities, where pharmacophore models or chemical similarity metrics can be adjusted to observe how these changes affect the ranking of potential hits. This interactivity significantly accelerates the iterative optimization process behind modern drug design. An example is Schrödinger LiveDesign integrating data from public and proprietary sources with interactive modeling tools that guide rational drug design while managing the complexity of structure-activity relationships.[51]

### 3.3. Systems Biology

Systems biology approaches benefit from interactive data environments that enable exploration of complex biological networks. Reactome and KEGG currently include interactive pathway browsers that enable to navigate from organism-level pathways down to molecular interactions, visualizing experimental data in context.[52] The ability to overlay multi-omics data (transcriptomics, proteomics, metabolomics) onto these pathways in real-time provides insights that would be impossible to extract from static representations. Advanced systems biology databases incorporate simulation capabilities, where researchers can interactively perturb network components and observe predicted system-wide effects. For example, Cell Collective allows users to build and simulate logical models of biological networks, interactively testing hypotheses about regulatory relationships.[53] These interactive modeling approaches bridge static pathway maps and dynamic biological processes to facilitate *in silico* experimentation.

### 3.4. Clinical Research

In clinical research, interactive data environments are transforming how patient cohorts are analyzed and stratified.[54] Modern clinical trial databases allow to dynamically segment patient populations based on multiple clinical variables, biomarkers, treatment responses, and genomic profiles. These systems enable the identification of responder subgroups that might be missed in traditional aggregate analyses. This interactive approach is particularly valuable for precision medicine, where treatment decisions increasingly depend on complex combinations of biomarkers. In this context, cBioPortal for Cancer Genomics is used by clinicians to interactively explore relationships between genomic alterations and clinical outcomes across thousands of patients, and identify patterns that inform treatment selection for individual cases.[54] As these systems evolve, they increasingly incorporate natural language processing of clinical notes and AI-assisted pattern recognition to extract insights from unstructured clinical data.

The applications discussed across genomics, drug development, systems biology, and clinical research demonstrate different approaches to biological data interactivity. To better illustrate the current limitations and development gaps, **Table 2** provides a systematic comparison of representative platforms in each domain, highlighting both implemented capabilities and features requiring further development.

Despite strong domain-specific performance, major limitations persist in cross-domain data integration. Systems must evolve to support standardized linking of genomic variants, pathway disruptions, drug targets, and clinical outcomes into unified, interactive analyses. In parallel, most platforms remain tailored for single-user interactions, lacking collaborative functions such as version control, permission management, or synchronized multi-user workspaces. Another limitation lies in the unidirectional flow of information. Current tools primarily serve as endpoints for querying existing knowledge rather than facilitating knowledge generation. Embedding structured annotation frameworks could enable users to contribute validated insights, fostering dynamic feedback loops between researchers and databases. Moreover, computational scalability remains a barrier: simple filtering and visualization are often responsive, but complex analyses are hindered by performance constraints. To overcome this, platforms will need distributed computing architectures optimized for biological data types. Finally, most systems offer only retrospective data exploration; integrating predictive modeling and simulation would enable hypothesis testing, allowing users to evaluate potential interventions prior to experimentation, significantly accelerating and improving research outcomes.

## 4. Discussion and Challenges

Despite their transformative potential, interactive data environments in life sciences face substantial challenges. The most intuitive challenges span from people unawareness, privacy and access, user experience, and intrinsically data-related challenges. People-related challenges revolve around the lack of awareness among part of life science researchers of what computing tools can offer to optimize, speed up, and improve the intellectual quality of life science research. This challenge reflects poor communication and limited exchange between life sciences and computational disciplines, which urgently need to be bridged. On the same line, user adoption represents a critical challenge. Interactive systems must accommodate diverse user groups with varying computational literacy while providing sufficient analytical depth to address complex biological questions.

Technical challenges are primarily about data standardization, performance, data quality, and accessibility. Data quality itself represents a central bottleneck, revolving around noisy or incomplete entries, inconsistent measurement protocols and experimental bias, which can critically impair the interpretability and reproducibility of downstream analyses. Sophisticated analyses that provide meaningful biological insights often require computational resources incompatible with real-time interaction. This creates a challenging design space where analytical depth must be balanced against performance constraints. Privacy and access challenges are particularly relevant in clinical and patient-derived datasets, involving ethical concerns, consent frameworks and jurisdictional restrictions. For example, the use and sharing of patient-level data must adhere to data protection regulations (e.g., GDPR, HIPAA), institutional review protocols, and evolving expectations around participant autonomy and trust. These constraints are essential for safeguarding rights and ethics, yet often introduce friction in integrating sensitive data across platforms. Initiatives such as the Global Alliance for Genomics and Health (GA4GH) are working to establish interoperable standards and policies to facilitate responsible data sharing while preserving privacy. Integrated data platforms intended to operate in this domain must be designed with embedded compliance layers and customizable permission systems.

## 4.1. Data Standardization

Biological data are produced by a wide variety of instruments and experimental methods, which often results in heterogeneous formats and varied quality. Standardizing data formats and ensuring data quality are fundamental challenges. Life science domains have developed specialized vocabularies that often overlap but use different terminologies for similar concepts. For example, cell lines may be described using inconsistent nomenclature across repositories (HeLa vs. HeLa S3 vs. Hela-S3), and the similarity across names could lead to misassignments. These misassignments, even if rare, could cause cascade problems. Ontology mapping has been already initiated, for example through OBO Foundry (Open Biological and Biomedical Ontologies) providing frameworks for ontology integration, but implementation remains challenging due to the evolving nature of biological knowledge.[55] Natural language processing models are increasingly employed to automatically map terms across vocabularies, but these systems require careful curation to validate mappings. The value of interactive queries depends fundamentally on the quality and completeness of underlying metadata, for instance experimental details necessary for proper interpretation. To address this challenge, interactive data environments could implement validation systems that flag missing critical data and provide feedback to contributors about data quality. Some systems now employ data provenance indicators or reputation scores for data sources, allowing users to filter query results based on source trustworthiness and metadata completeness. Examples are BioThings Explorer and FAIRsharing.[56, 57]

## 4.2. Performance

As databases grow in size and complexity, ensuring that query responses remain fast and accurate is crucial. Interactive medical biotechnology queries involve complex multidimensional parameters. For example, identification of cell lines with specific genetic mutations, protein expression patterns, and growth characteristics, alongside dynamic visualization of multiscale relationships across tens of cell types and culture. Advanced computational approaches addressing this challenge include bitmap indexing for genomic data and spatial indexing techniques adapted for multidimensional biological data.[58]

Cloud-native database architectures that scale horizontally to handle compute-intensive queries are increasingly essential for interactive performance across large biological datasets. Interactive visualization of large biological datasets presents unique performance challenges. Modern interactive data environments address this through server-side aggregation, progressive loading techniques that refine visualizations incrementally, WebGL and GPU-accelerated rendering, and intelligent sampling methods that preserve statistical properties while reducing data volume.[59] These techniques enable responsive exploration even for datasets too large to transmit in their entirety. This transition can be enabled step by step, handling datasets that are easier to manage first.

## 4.3. Data Accessibility

Advanced interactivity can only be effective if users find the system intuitive and accessible. Interactive biological databases face a fundamental issue between analytical power and interface simplicity. Systems that expose the full complexity of underlying data models risk overwhelming non-computational users, while oversimplified interfaces may limit discovery potential. This challenge is particularly acute in multidisciplinary fields where users range from computational specialists to wet-lab biologists and clinicians. Adaptive interface approaches show promise in addressing this challenge through progressive disclosure of features based on user expertise, context-sensitive guidance, customizable workspaces, and natural language query capabilities for nontechnical users. Features such as automatic query history tracking (e.g., as implemented in NCBI resources) and computational notebooks (e.g., Jupyter) are now widely adopted tools that have proven effective in enhancing reproducibility, transparency, and user engagement when integrated into interactive database interfaces. Another issue related to data accessibility concerns the integration and retrieval of information from existing biological databases, which is often hindered by inconsistent formats, limited APIs, or restricted access policies.

## 5. Outlook

interactive data environments could represent a paradigm shift in how research data is stored, handle, and shared, offering significant advantages toward driving collective scientific progress meant for clinical translation and reliable fundamental results. Their emergence signals not merely a technological evolution but a fundamental shift in how biological knowledge is constructed, validated, and extended. This reconceptualizes the scientific process itself where the boundaries between hypothesis generation, data analysis, and experimental design become increasingly iterative, with a strong urge for data reproducibility and validation. Currently, different expertise in computational methods creates a gap between those who can and cannot effectively interrogate complex biological datasets. Integrated frameworks designed with intuitive interfaces could democratize access to advanced analytical capabilities, potentially shifting control over data interpretation and exploratory analysis from computational specialists to a broader range of scientists. The development trajectory of interactive data environments will inevitably be shaped by economic forces and institutional priorities that extend beyond purely scientific considerations. Commercial entities building such platforms face tensions between creating proprietary systems that generate revenue and contributing to open scientific models that maximize knowledge generation and sharing. These economic realities suggest that hybrid models combining open source models with commercial components may emerge.

Interactive data environments also have the potential to transform interdisciplinary collaboration by creating shared cognitive spaces where specialists from diverse backgrounds can explore complex biological questions. This potential extends beyond collaboration among human experts to include the integration of AI-driven tools that support data exploration and interpretation, while preserving human oversight and decision-making. Static repositories indirectly reinforce reductionist perspectives by presenting biological entities as discrete objects. Interactive systems that dynamically visualize multidimensional relationships could instead represent the existing interconnections between the most seemingly disparate domains, making full use of the acquired data across domains.

## Conflicts of Interest

The author declares no conflicts of interest.

## List of tables

**Table 1. Comparison between traditional biological databases and next-generation interactive data environments.** This table summarizes key architectural, functional, and epistemological differences between classical repositories, designed primarily for data storage and retrieval, and the proposed integrated systems, which emphasize bidirectional data flow, real-time analytics, collaborative workflows, and integration of experimental design logic and user feedback.

| Dimension | Classical biological databases | Interactive data environments |
|---|---|---|
| Access model | Read-only, query-based | Bidirectional and real-time |
| Data structure | Schema-defined | Flexible (relational, document, graph) |
| User engagement | Individual | Multi-user |
| Update frequency | Curated (low frequency) | Real-time ingestion & user feedback |
| Knowledge generation | Initiated by the user | Embedded in dynamic workflows |
| Feedback | Limited to curation | Real-time FAIR integration |
| Use case focus | Archival and citation | Discovery and planning |

**Table 2. Comparison of existing integrated data environments across major medical biotechnology domains.** The table summarizes implemented features and highlights missing capabilities across representative platforms in genomics, drug development, systems biology, and clinical research, in relation to the proposed interactive model.

| Category | Feature | Genomics (gnomAD) | Drug development (DrugBank) | Systems biology (Reactome) | Clinical research (cBioPortal) |
|---|---|---|---|---|---|
| **Data structure and storage** | Multi-omics data | Variant-focused with limited transcriptomic data | Chemical-biological integration | Pathway-focused with limited multi-omics | Multi-omics with clinical data |
| | Unstructured data | Limited text mining | Chemical literature | Protocol documentation | Clinical notes processing |
| | Temporal data support | Static datasets | Limited reaction kinetics | Dynamic pathway simulation | Longitudinal patient data |
| **Query and visualization** | Real-time filtering of $>10^6$ records | Population-scale variant filtering | Limited to subsets of compound database | Pathway-limited scope | Sample-limited cohort selection |
| | Interactive cross-domain queries | Limited to genomic context | Drug-target-pathway connections | Within pathway boundaries | Genotype-phenotype correlations |
| | 3D/spatial data visualization | Limited protein structure | Molecular structure viewers | Network topology | Limited anatomical context |
| **Analytics and interactivity** | Interactive statistical analysis | Population frequency tools | Structure-activity analysis | Enrichment analysis | Survival and correlation analysis |
| | Hypothesis testing framework | Limited to constraint metrics | Virtual screening | Pathway perturbation | Biomarker association testing |
| | Machine learning integration | Variant pathogenicity | Chemical similarity | Limited predictive models | Outcome prediction |
| **Collaborative features** | Multi-user simultaneous interaction | Single-user model | Single-user model | Single-user model | Limited sharing capabilities |
| | Version control of analyses | Download-only | Limited project saving | Export options | Study groups |
| | User annotation frameworks | None | Limited annotations | None | Basic study descriptions |
| **Technical architecture** | API extensibility | Comprehensive REST API | Basic REST endpoints | Limited API access | Comprehensive programmatic access |
| | Computational scalability | Cloud-based distributed computing | Limited analytical capacity | Server-based processing | Hybrid architecture |
| | FAIR data principles implementation | Partial implementation | Partial implementation | Partial implementation | Partial implementation |

## List of figures

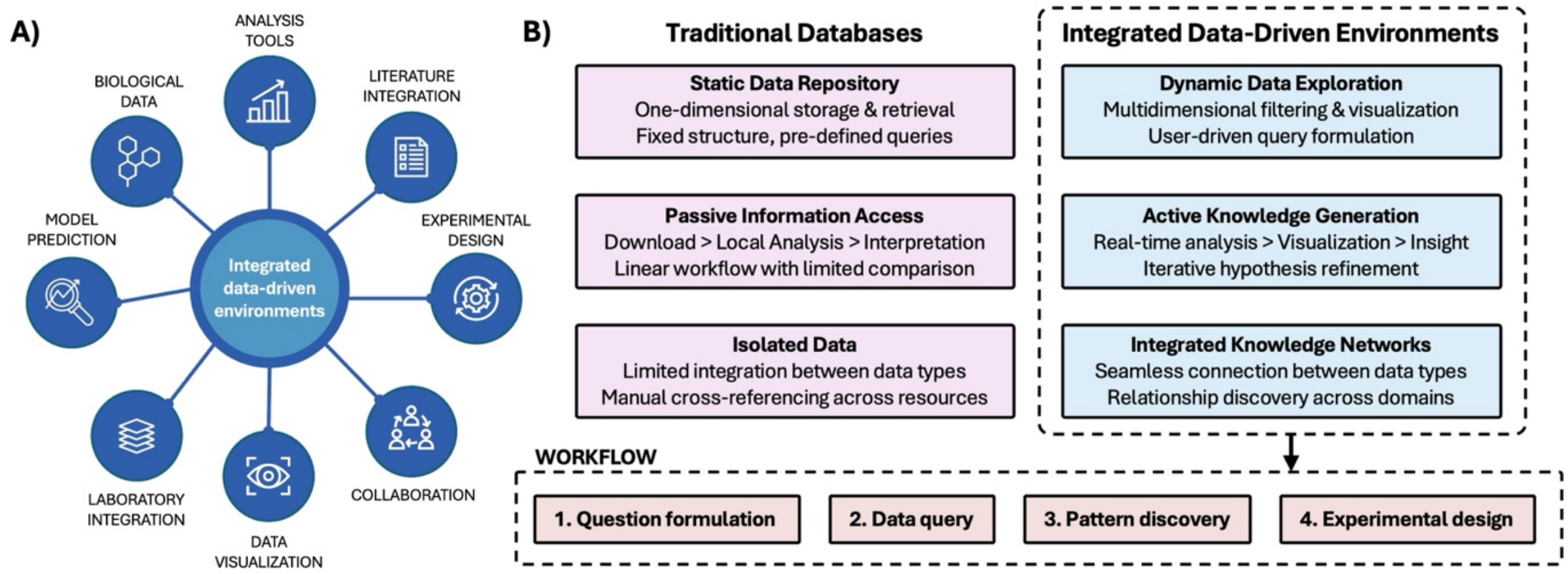

**Figure 1. Integrated data environments.** The image at the left displays the concept of integrated data environments, needs and features. The diagram at the right compares static repositories with interactive platforms across three key dimensions: data access methods, analysis workflows, and knowledge integration capabilities. The workflow at the bottom exemplifies the steps undertook by the user interfacing with an interactive data environment

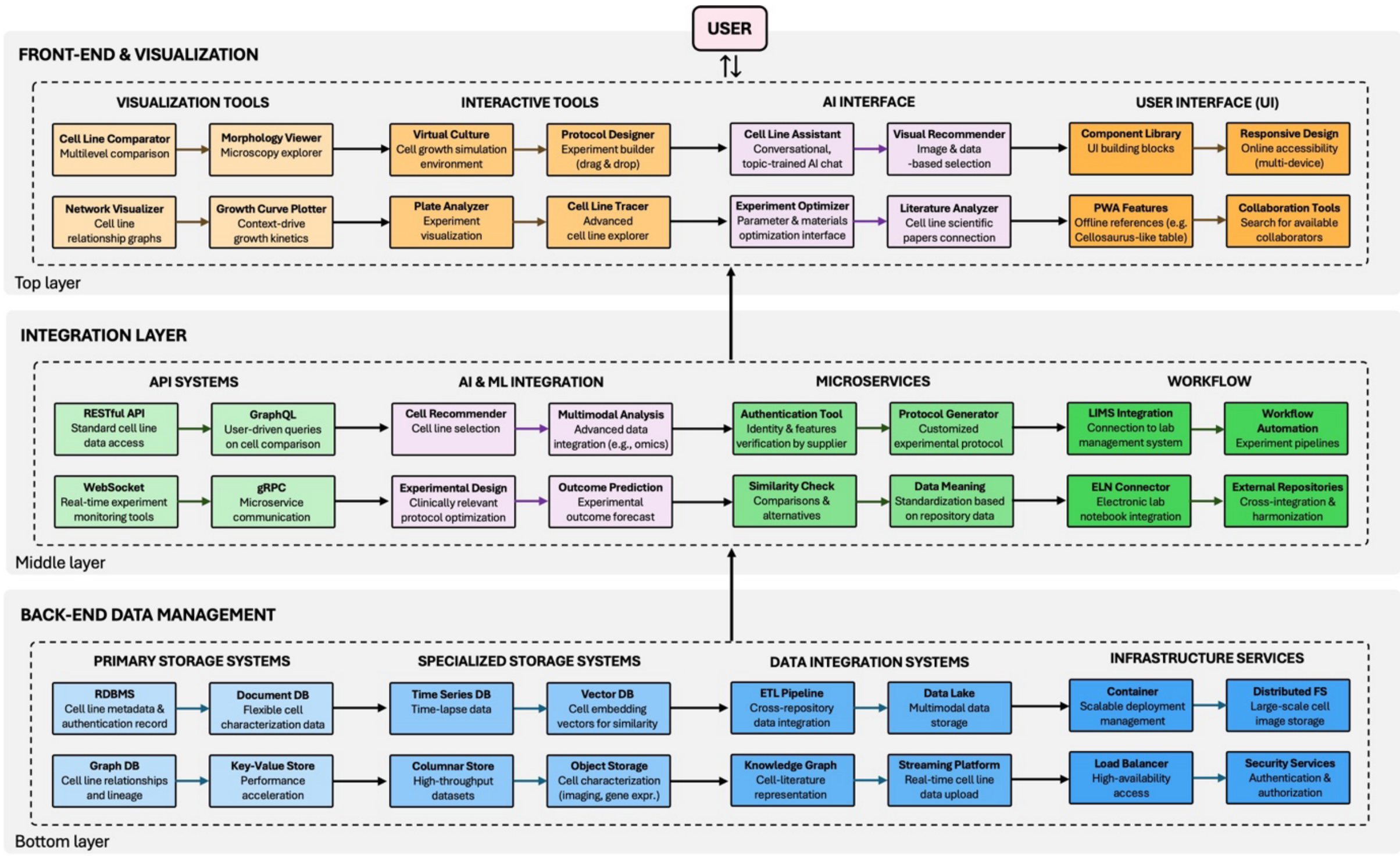

**Figure 2. Technical architecture of integrated data environments: a case study on informed cell line selection.** The schematic illustrates the three-tier structure comprising back-end data management systems (databases and storage solutions), middleware integration layer (APIs, microservices, AI components), and front-end technologies (visualization tools and user interfaces), tailored to cell line selection as a sample case.